\documentclass[prb,showpacs,twocolumn,aps,superscriptaddress,floatfix]{revtex4}
\usepackage{amsmath}
\usepackage{amssymb}
\usepackage{bm}
\usepackage{graphicx}
\usepackage{color}
\usepackage{ulem}
\usepackage{hyperref}
\usepackage[T1]{fontenc}

\begin{document}

\title{
Single-electron gap in the spectrum of twisted bilayer graphene}

\author{A.V. Rozhkov}
\affiliation{CEMS, RIKEN, Wako-shi, Saitama, 351-0198, Japan}
\affiliation{Institute for Theoretical and Applied Electrodynamics, Russian Academy of Sciences, 125412 Moscow, Russia}
\affiliation{Moscow Institute of Physics and Technology, Dolgoprudny,
Moscow Region, 141700 Russia}

\author{A.O. Sboychakov}
\affiliation{CEMS, RIKEN, Wako-shi, Saitama, 351-0198, Japan}
\affiliation{Institute for Theoretical and Applied Electrodynamics, Russian Academy of Sciences, 125412 Moscow, Russia}

\author{A.L. Rakhmanov}
\affiliation{CEMS, RIKEN, Wako-shi, Saitama,
351-0198, Japan}
\affiliation{Institute for Theoretical and Applied Electrodynamics, Russian Academy of Sciences, 125412 Moscow, Russia}
\affiliation{Moscow Institute of Physics and Technology, Dolgoprudny,
Moscow Region, 141700 Russia}
\affiliation{All-Russia Research Institute of Automatics, Moscow, 127055
Russia }

\author{Franco Nori}
\affiliation{CEMS, RIKEN, Wako-shi, Saitama, 351-0198, Japan}
\affiliation{Department of Physics, University of Michigan, Ann Arbor, MI
48109-1040, USA}

\begin{abstract}
We investigate the gap in the single-electron spectrum of twisted bilayer
graphene. In a perfect infinite lattice of a twisted bilayer, the gap
varies exponentially in response to weak changes of the twist angle. Such a large
sensitivity makes theoretical predictions of the gap nearly impossible,
since experimentally the twist angle is always known with finite accuracy.
To address this issue, we numerically study finite clusters of twisted bilayer graphene. For finite systems, changing the twist angle causes
a gradual crossover between gapless and gapped regimes. The crossover
occurs when the finite-size quantization energy becomes comparable to the
matrix elements responsible for the generation of the gap. We further argue
that disorder scattering can induce similar crossover, in which the
mean-free path plays the same role as the system size for the finite
clusters. It is demonstrated that, to observe the gap experimentally, it is
necessary to have a sample of suitable purity, and to possess the ability
to tune the twist angle accurately.
\end{abstract}


%73.21.-b 	Electron states and collective excitations in multilayers,
%quantum wells, mesoscopic, and nanoscale systems (for electron states in
%nanoscale materials, see 73.22.-f)
%
%73.21.Ac 	Multilayers
%
%73.21.Cd 	Superlattices
%
%73.21.Fg 	Quantum wells
%
%73.21.Hb 	Quantum wires
%
%73.21.La 	Quantum dots
%
%73.22.-f 	Electronic structure of nanoscale materials and related
%systems
%
%73.22.Dj 	Single particle states
%
%73.22.Gk 	Broken symmetry phases
%
%73.22.Lp 	Collective excitations
%
%73.22.Pr 	Electronic structure of graphene

\maketitle

\section{Introduction}

Recent experimental studies (scanning tunneling microscopy,
STM~\cite{STM2,DF_TEM1,STM1,STM_VHS2}, Raman
spectroscopy~\cite{NiRamanVrenorm,DF_TEM2},
angular resolved photoemission
spectroscopy~\cite{OhtaARPES,ARPES_NatMat})
revealed that, in many cases, the structure of bilayer graphene samples is
far from the ideal AB stacking. Instead, it is characterized by a non-zero
twist angle $\theta$ between graphene layers. The electronic structure of
twisted bilayer graphene (tBLG) is very rich, demonstrating a Dirac
spectrum with a $\theta$-dependent Fermi
velocity~\cite{STM2,NiRamanVrenorm},
low-energy van Hove
singularities~\cite{STM1,STM_VHS2},
complex Fermi surface~\cite{ourTBLG,Pankratov5},
and other peculiar
features~\cite{LeeQHE,tBLG_QHE}.
An important characteristic of its electronic structure is the
single-electron gap. For twisted bilayer samples, the existence of the gap
was demonstrated in several
experiments~\cite{ARPES_NatMat,TransportGap2015}.
This paper theoretically studies the gap (previous efforts on this issue
are discussed in the recent review paper in
Ref.~\onlinecite{ourReview}).

If one is interested in the theoretical description of the tBLG, a useful
starting point is to consider `commensurate' values of $\theta$ for which
the tBLG lattice forms commensurate superstructures. When the size of the
supercell is not too large, the electronic properties can be studied
numerically~\cite{kolmogorov,LatilDFT,PankratovPRB2013,Pankratov1,
PankratovFlakes,PankratovPRL,NanoLettTB,TramblyTB_Loc,Morell1,ourTBLG,
Pankratov5,Peeters2016}.
Besides computational approaches, several semi-analytic theories for
low-energy electrons were
developed~\cite{MeleReview,dSPRL,dSPRB,PNAS,MelePRB1,MelePRB2,
NonAbelianGaugePot}.
Studying the commensurate angles, it is possible to calculate, for example,
the dependence on $\theta$ of the Fermi
velocity~\cite{dSPRL,dSPRB,PNAS}
$v_{\rm F}$
and the density of
states~\cite{ourTBLG}.
Unfortunately, these
approaches cannot be directly applied for the calculation of the gap. It
was demonstrated in
Ref.~\onlinecite{ourTBLG}
that the gap $\Delta$ evaluated at the commensurate angles is not a smooth
function of $\theta$. Instead, it varies exponentially even for small
changes of the twist angle. Clearly, such a large sensitivity implies that
considering the commensurate angles is not sufficient for a
consistent theory of how the gap is generated.

A possible way to remedy this situation was proposed in
Ref.~\onlinecite{ourTBLG}.
It was pointed out that the sharp jumps of $\Delta$ were associated with
the fact that the size of the supercell may change drastically for very
small variations of $\theta$. Therefore, the extreme sensitivity of $\Delta$
to the twist angle is possible only in a perfect infinite lattice of tBLG,
where a superstructure with arbitrary large supercell can exist. Of course,
any real sample has a finite linear size $L$. Furthermore, a realistic
electron propagation is characterized by a finite mean free path
$l_{\rm m}$
due to electron scattering on defects, such as impurities, ``wrinkles'' (as an example, below we will evaluate
$l_{\rm m}$
for a particular case of a disordered ensemble of one-dimensional
``wrinkles''), etc. The smallest among the length scales $L$ and
$l_{\rm m}$
would introduce a ``cutoff'', which disallows the superstructures with
large supercells, and makes the jumps of $\Delta$
impossible~\cite{ourTBLG}.

The latter reasoning motivates us to investigate the formation of the gap
in a tBLG sample of finite size. For tBLG clusters of various twist angles
and linear sizes, we numerically determine the matrix elements, which
couple different Dirac cones. By construction, the calculated matrix
elements are smooth functions of $\theta$. Since these matrix elements are
small in comparison to the graphene band-width, many publications often
dismiss them. Yet, they are important at low energies, causing qualitative
changes to the electron spectrum: in the ideal infinite tBLG lattice they
either open the gap, or induce a so-called ``band splitting''. In a
finite-size sample, or in a sample with finite quasiparticle scattering,
these cone-coupling matrix elements require a subtler interpretation: a gap
cannot be observed, unless the corresponding matrix element exceeds both
the dimensional quantization gap, and quasiparticle scattering frequency.
We will demonstrate that this condition is satisfied only when $\theta$ is
close to a commensurate angle with small supercell size. As the detuning
from the ``good'' angle increases, the gap-generating matrix elements
quickly (exponentially) decay, and the gap is washed away by the external
scattering.

The paper is organized as follows.
Section~\ref{sect::geometry}
summarizes the geometry of the tBLG lattice. In
Sec.~\ref{sect::prelim}
we discuss the general theoretical background of the problem considered.
The scattering on the linear defects (``wrinkles''), which is a very
effective mechanism limiting the coherent propagation of the electrons in
graphene, is studied in
Sec.~\ref{sect::scattering}.
The numerical results for the finite-size samples are presented in
Sec.~\ref{sect::numerics}.
The discussion and conclusions are given in
Sec.~\ref{sect::disc_and_concl}. Additional details of the calculation of the matrix elements are presented in the Appendix.

\section{Geometry of twisted bilayer lattice}
%%%%%%%%%%%%%%%%%%%%%%%%%%%%%%%%%%%%%%%%%%%%%%%%%%
\label{sect::geometry}
%%%%%%%%%%%%%%%%%%%%%%%%%%%%%%%%%%%%%%%%%%%%%%%%%%

In this section, for reader's convenience, we provide basic information
about the geometry of the twisted bilayer lattice. This will allow to
introduce equations and notation which will be used later throughout this
paper. The presentation here follows
Refs.~\onlinecite{ourTBLG,ourReview}. A more general and comprehensive consideration of the slightly mismatched overlayers is done  in Ref.~\onlinecite{Hermann}.

A bilayer consists of two layers, one lying over the other. We will assume
that the layers are perfectly flat, and separated by the distance
$d=3.35$\,{\AA}
from each other. In a real tBLG sample the layers are not purely
two-dimensional. The interlayer distance
varies~\cite{STM1}
depending on the local arrangement of the atoms. However, the interlayer
corrugation is quite small
($\sim 0.1$\,\AA),
and our approximation is well-justified.

\begin{figure}[t]
\centering
\includegraphics[width=0.8\columnwidth]{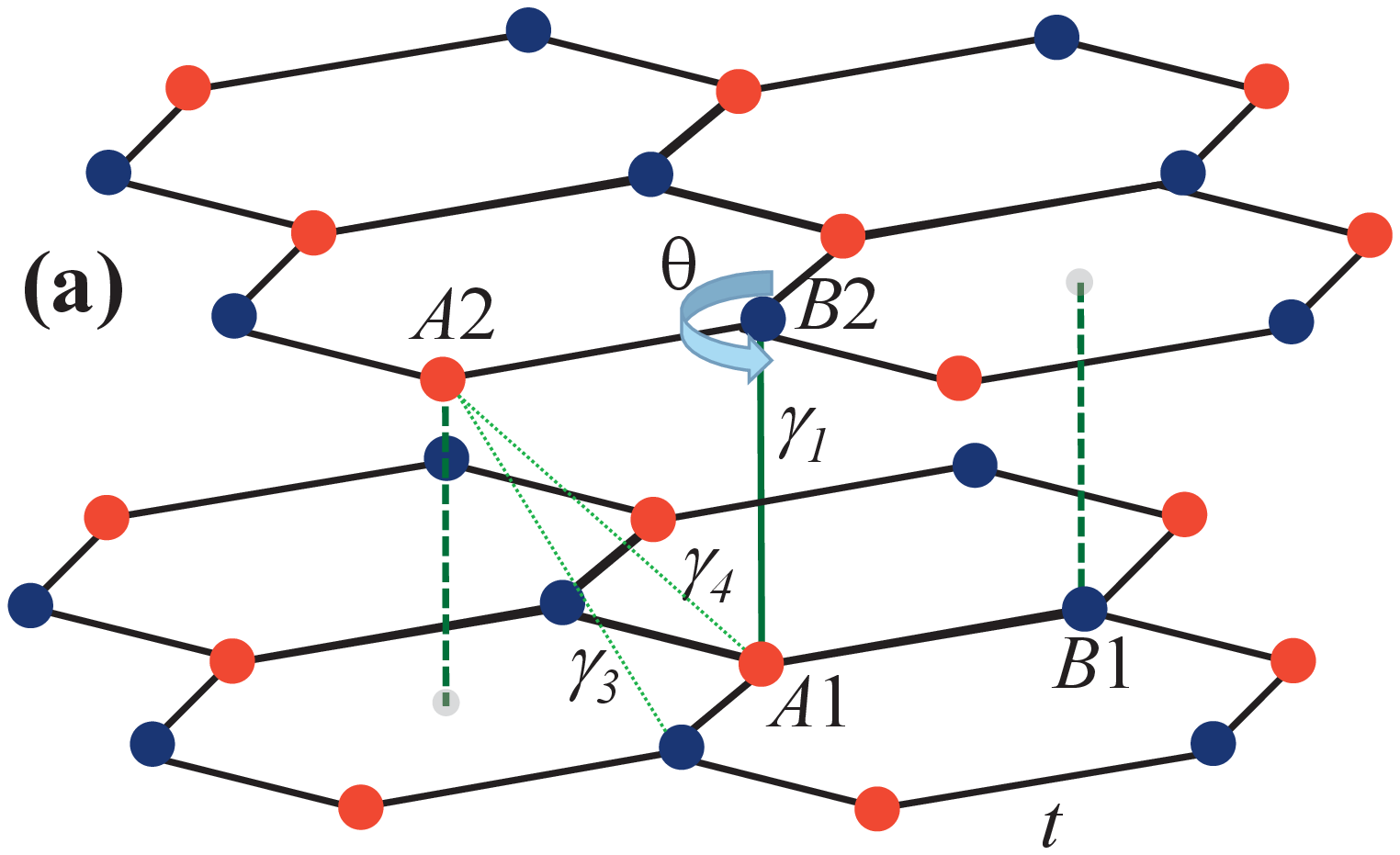}\\
\includegraphics[width=0.8\columnwidth]{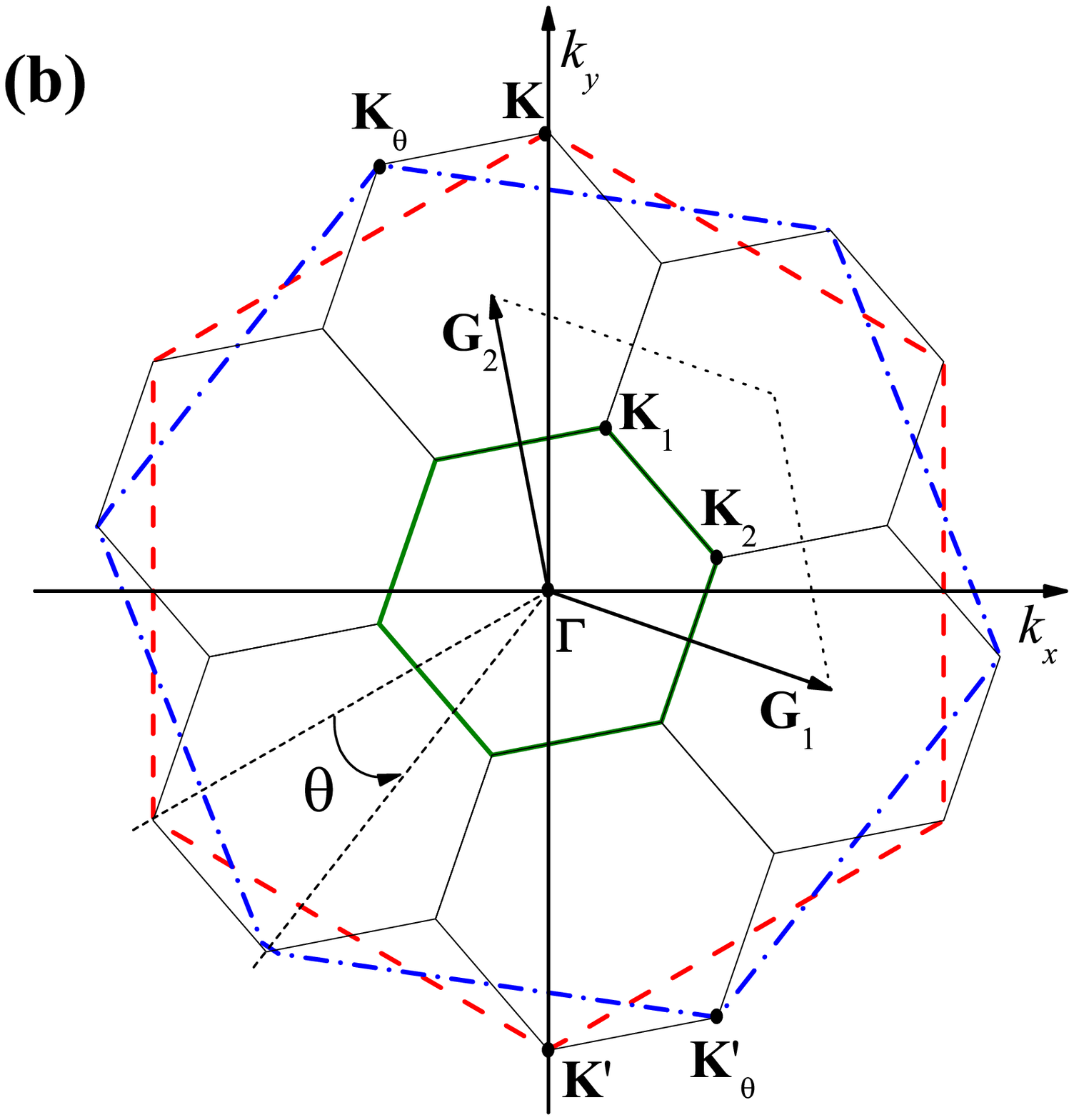}
\caption{(a) Starting from a perfect AB-bilayer graphene, a twisted
graphene bilayer is obtained by rotating the top layer by the angle
$\theta$ (shown by the blue rotating arrow). The rotation is performed
around the axis connecting sites
$A1$
and
$B2$;
the quantity $t$ is the in-plane nearest-neighbor hopping, and
$\gamma_{1,3,4}$,
are out-of-plane hopping amplitudes of the AB-stacked bilayer. These
$\gamma$s are used to fix the fitting parameters of the function
$t_{\bot}(\mathbf{r};\mathbf{r}')$
(see the text). In this paper we use
$\gamma_1 = 0.4$\;eV,
$\gamma_3 = 0.254$\;eV, and
$\gamma_4 = 0.051$\;eV,
which are all substantially smaller than the in-plane hopping amplitude
$t = 2.8$\;eV.
(b) Reciprocal space structure for
$\theta=21.787^{\circ}$
($m_0=1$,
$r=1$).
The large hexagons show the Brillouin zones of individual layers: the red
dashed (the blue dot-dashed) hexagon corresponds to the bottom (top) layer.
The green thick solid hexagon represents the first Brillouin zone of the
bilayer. The next several Brillouin zones of the tBLG are shown by black
thin solid hexagons. The Dirac point
$\mathbf{K}'$
($\mathbf{K}'_{\theta}$)
is equivalent to the point
$\mathbf{K}_{\theta}$
($\mathbf{K}$)
if
$r\neq3n$.
When
$r=3n$,
$\mathbf{K}_{\theta}\sim\mathbf{K}$
and
$\mathbf{K}'_{\theta}\sim\mathbf{K}'$
(see the text). The tBLG Dirac points
$\mathbf{K}_{1,2}$
are doubly degenerate: each of them is equivalent to one of two Dirac
points of each graphene layer. For the particular case of the
$(1,1)$
superstructure,
$\mathbf{K}_1\sim\mathbf{K}\sim\mathbf{K}'_{\theta}$,
and
$\mathbf{K}_2\sim\mathbf{K}'\sim\mathbf{K}_{\theta}$.
}
%%%%%%%%%%%%%%%%%%%%%%%%%%%%%%%%%%%%%%%%%%%%%%%%%%
\label{FigTBLG}
%%%%%%%%%%%%%%%%%%%%%%%%%%%%%%%%%%%%%%%%%%%%%%%%%%
\end{figure}

Each graphene layer consists of two sublattices,
$A1$
and
$B1$
in the layer~1 (bottom layer, see
Fig.~\ref{FigTBLG}a),
and
$A2$,
$B2$
in the layer~2 (top layer). In the layer~1 the positions of the carbon
atoms are given by the equations
\begin{eqnarray}
%%%%%%%%%%%%%%%%%%%%%%%%%%%%%%%%%%%%%%%%%%%%%%%%%%
\label{eq::lattice1}
%%%%%%%%%%%%%%%%%%%%%%%%%%%%%%%%%%%%%%%%%%%%%%%%%%
\mathbf{r}_{\mathbf{n}}^{1A}\equiv\mathbf{r}_{\mathbf{n}}
=
n\mathbf{a}_1 + m\mathbf{a}_2\,,\;
\mathbf{r}_{\mathbf{n}}^{1B}
=
\mathbf{r}_{\mathbf{n}} + \bm{\delta}_1\,,\\[5pt]
\bm{\delta}_1
=
\frac{1}{3}(\mathbf{a}_1+\mathbf{a}_2)
=
a(1/\sqrt{3},\,0)\,,
\end{eqnarray}
where
$\mathbf{n}=(n,m)$
is a vector with integer-valued components $n$ and $m$, the vector
$\bm{\delta}_1$
points to a nearest-neighbor site on the honeycomb lattice, and
$\mathbf{a}_{1,2}$
are primitive vectors of the lattice
\begin{equation}
\mathbf{a}_1=\frac{a}{2}(\sqrt{3},\,-1),
\qquad
\mathbf{a}_2=\frac{a}{2}(\sqrt{3},\,1),
\end{equation}
with the lattice parameter
$a=2.46$\,{\AA}.
We will also use the length of the in-plane carbon-carbon bond
$a_0 = a/\sqrt{3} = 1.42$\,{\AA}.

When
$\theta = 0$,
the system is a perfect AB bilayer. Let us consider the situation when the
layer~2 is rotated with respect to layer~1 by the angle $\theta$ around the
axis connecting the atoms
$A1$
and
$B2$
with
$\mathbf{n}=0$
(see
Fig.~\ref{FigTBLG}).
The atoms of the rotated layer, thus, have the positions
\begin{eqnarray}
%%%%%%%%%%%%%%%%%%%%%%%%%%%%%%%%%%%%%%%%%%%%%%%%%%
\label{eq::lattice2}
%%%%%%%%%%%%%%%%%%%%%%%%%%%%%%%%%%%%%%%%%%%%%%%%%%
\mathbf{r}_{\mathbf{n}}^{2B}
\equiv
\mathbf{r}'_{\mathbf{n}}
=
n\mathbf{a}'_1 + m\mathbf{a}'_2\,,\;
\mathbf{r}_{\mathbf{n}}^{2A}
=
\mathbf{r}'_{\mathbf{n}}-\bm{\delta}'_1\,,
\end{eqnarray}
where
\begin{eqnarray}
\mathbf{a}'_{1,2}=\mathbf{a}_{1,2}
\left(	
	\cos\theta\mp\frac{\sin\theta}{\sqrt{3}}
\right)
\pm
\mathbf{a}_{2,1}\frac{2\sin\theta}{\sqrt{3}}\,,\\
\bm{\delta}'_1=\frac{a}{\sqrt{3}}(\cos\theta,\, \sin\theta)\,.
\end{eqnarray}

The structure of the tBLG is commensurate
if~\cite{dSPRL,dSPRB,PankratovPRB2013,MeleReview}
\begin{equation}\label{theta}
\cos\theta=\frac{3m_0^2+3m_0r+r^2/2}{3m_0^2+3m_0r+r^2}\,,
\end{equation}
where
$m_0$
and
$r$
are coprime positive integers. For these angles the superlattice vectors
$\mathbf{R}_{1,2}$
are:
\begin{equation}
\left\{
\begin{array}{rcl}
\mathbf{R}_1&=&m_0\mathbf{a}_1+(m_0+r)\mathbf{a}_2\\
\mathbf{R}_2&=&-(m_0+r)\mathbf{a}_1+(2m_0+r)\mathbf{a}_2
\end{array}\right.\,(r\neq3n,\;n\in\mathbb{N}),
\end{equation}
or
\begin{equation}
\left\{
\begin{array}{rcl}
\mathbf{R}_1&=&(m_0+n)\mathbf{a}_1+n\mathbf{a}_2\\
\mathbf{R}_2&=&-n\mathbf{a}_1+(m_0+2n)\mathbf{a}_2
\end{array}\right.\,(r=3n,\;n\in\mathbb{N}).
\end{equation}

An important property of the superlattice is the number of sites in a
supercell. It equals to
\begin{equation}
\label{Nsc}
N(m_0,r)=\left\{\begin{array}{l}
4(3m_0^2+3m_0r+r^2),\;\text{if}\;r\neq3n\,,\\
4(m_0^2+m_0r+r^2/3),\;\text{if}\;r=3n\,.
\end{array}\right.
\end{equation}
The linear size of the superlattice cell is
$L_{\rm sc}\equiv|\mathbf{R}_{1,2}|=a\sqrt{N}/2$.

The primitive vectors of the reciprocal superlattice can be written as
\begin{eqnarray}
\mathbf{G}_1
&=&
\frac{(2m_0+r)\mathbf{b}_1+(m_0+r)\mathbf{b}_2}{3m_0^2+3m_0r+r^2}\,,
\nonumber
\\
\mathbf{G}_2
&=&
\frac{-(m_0+r)\mathbf{b}_1+m_0\mathbf{b}_2}{3m_0^2+3m_0r+r^2},\quad \textrm{if}\,\, r\neq3n,
\end{eqnarray}
or
\begin{eqnarray}
\mathbf{G}_1&=&\frac{(m_0+2n)\mathbf{b}_1+n\mathbf{b}_2}{m_0^2+m_0r+r^2/3}\,,\nonumber\\
\mathbf{G}_2&=&\frac{-n\mathbf{b}_1+(m_0+n)\mathbf{b}_2}{m_0^2+m_0r+r^2/3},\quad \textrm{if}\,\, r=3n,
\end{eqnarray}
where
$\mathbf{b}_{1,2}$
are the reciprocal lattice vectors of the single layer graphene
\begin{equation}
\mathbf{b}_1
=
\frac{2\pi}{\sqrt{3} a}(1,\,-\sqrt{3})\,,
\qquad
\mathbf{b}_2=\frac{2\pi}{\sqrt{3} a}(1,\,\sqrt{3})\,.
\end{equation}
The first Brillouin zone of the superlattice has the shape of a hexagon
with side
$|\mathbf{G}_{2}-\mathbf{G}_{1}|/3$.
In the particular case
$r=1$,
this side is equal to
$\Delta K=|\mathbf{K}_{\theta}-\mathbf{K}|$,
where
\begin{eqnarray}
\mathbf{K}=\frac{4\pi}{3a}(0,\,1)
\quad
\text{and}
\quad
\mathbf{K}_{\theta}=\frac{4\pi}{3a}(-\sin\theta,\,\cos\theta)
\end{eqnarray}
are the Dirac points of the bottom and top layers, respectively. The
electron states near the points
$\mathbf{K}$
and
$\mathbf{K}_{\theta}$
have identical chiralities. The points of opposite chirality are located at
$\mathbf{K}'=-\mathbf{K}$
and
$\mathbf{K}'_{\theta}=-\mathbf{K}_{\theta}$.
In the Brillouin zone of the superstructure, the Dirac points coordinates are given by the following expressions
\begin{eqnarray}
\mathbf{K}&=&-\mathbf{K}'=m_0\mathbf{G}_2+\frac{r}{3}\left(\mathbf{G}_1+2\mathbf{G}_2\right),\nonumber\\
\mathbf{K}_{\theta}&=&-\mathbf{K}'_{\theta}=m_0\mathbf{G}_2+\frac{r}{3}\left(\mathbf{G}_2-\mathbf{G}_1\right),\label{Kn3n}
\end{eqnarray}
if
$r\neq3n$,
or
\begin{eqnarray}
\mathbf{K}&=&-\mathbf{K}'=\frac{r}{3}\mathbf{G}_2+\frac{m_0}{3}\left(\mathbf{G}_2-\mathbf{G}_1\right),\nonumber\\
\mathbf{K}_{\theta}&=&-\mathbf{K}'_{\theta}=-\frac{r}{3}\mathbf{G}_1+\frac{m_0}{3}\left(\mathbf{G}_2-\mathbf{G}_1\right),
%%%%%%%%%%%%%%%%%%%%%%%%%%%%%%%%%%%%%%%%%%%%%%%%%%
\label{K3n}
%%%%%%%%%%%%%%%%%%%%%%%%%%%%%%%%%%%%%%%%%%%%%%%%%%
\end{eqnarray}
if
$r=3n$.

One can check that, if
$r\neq3n$,
point
$\mathbf{K}'$
is equivalent to
$\mathbf{K}_{\theta}$,
and
$\mathbf{K}$
is equivalent to
$\mathbf{K}'_{\theta}$:
\begin{eqnarray}
%%%%%%%%%%%%%%%%%%%%%%%%%%%%%%%%%%%%%%%%%%%%%%%%%%
\label{eq::class1}
%%%%%%%%%%%%%%%%%%%%%%%%%%%%%%%%%%%%%%%%%%%%%%%%%%
\mathbf{K}' \sim \mathbf{K}_{\theta}
\quad
\text{and}
\quad
\mathbf{K} \sim \mathbf{K}'_{\theta}
\quad
\text{for\ \ }r\neq3n.
\end{eqnarray}
Indeed, for such a value of
$r$,
the difference
$\mathbf{K}'-\mathbf{K}_{\theta}$
is a reciprocal vector of the superlattice. When
$r=3n$,
the equivalency relations are different:
\begin{eqnarray}
%%%%%%%%%%%%%%%%%%%%%%%%%%%%%%%%%%%%%%%%%%%%%%%%%%
\label{eq::class2}
%%%%%%%%%%%%%%%%%%%%%%%%%%%%%%%%%%%%%%%%%%%%%%%%%%
\mathbf{K}\sim\mathbf{K}_{\theta}
\quad
\text{ and }
\quad
\mathbf{K}'\sim\mathbf{K}'_{\theta}
\quad
\text{for\ \ }r=3n.
\end{eqnarray}
Thus, for any commensurate angle we have two doubly-degenerate
non-equivalent Dirac points of the tBLG. It follows from
Eqs.~\eqref{Kn3n}
and~\eqref{K3n}
that inside the reciprocal cell of the superlattice, the two non-equivalent
tBLG Dirac points are located at
\begin{equation}\label{K12}
\mathbf{K}_1=\frac{\mathbf{G}_1+2\mathbf{G}_2}{3}\,,\;\;\mathbf{K}_2=\frac{2\mathbf{G}_1+\mathbf{G}_2}{3}\,.
\end{equation}
Double degeneracy of these Dirac cones affects the electronic structure of
the tBLG leading to the band splitting and band gap formation.

Besides
$L_{\rm sc}$,
the tBLG has another characteristic length scale. The rotation of one
graphene layer with respect to another leads to the appearance of Moir\'{e}
patterns, manifesting in STM
experiments~\cite{STM2,DF_TEM1,STM1,STM_VHS2}
as alternating bright and dark regions. The Moir\'{e} period
$L_{\rm M}$
is defined as the distance between the centers of two neighboring bright
(or dark) regions. It is related to the twist angle as
\begin{equation}
%%%%%%%%%%%%%%%%%%%%%%%%%%%%%%%%%%%%%%%%%%%%%%%%%%
\label{MoireP}
%%%%%%%%%%%%%%%%%%%%%%%%%%%%%%%%%%%%%%%%%%%%%%%%%%
L_{\rm M}=\frac{a}{2\sin(\theta/2)}\,.
\end{equation}
It is possible to establish that the superstructure coincides with the
Moir\'{e} pattern when
$r=1$.
For other superstructures,
$L_{\rm sc}$
is greater than
$L_{\rm M}$.
The supercells of these structures contain
$r^2$
(if
$r\neq3n$)
or
$r^2/3$
(if
$r=3n$)
Moir\'{e} cells, and the arrangements of atoms inside these Moir\'{e} cells
are slightly different from each other. This means, in particular, that the
structures with
$r>1$
can be considered as almost periodic
repetitions~\cite{dSPRB}
of structures with
$r=1$.
The Moir\'{e} pattern and the superstructure are two complementary concepts
used to describe the tBLG.

The Moir\'{e} pattern depends smoothly on the twist angle, as demonstrated
by
Eq.~(\ref{MoireP}),
and can be easily detected experimentally. However, working with the
Moir\'{e} theoretically may be challenging since the Moir\'{e} structure is
strictly periodic for a very limited discrete set of angles. For a generic
value of $\theta$, different Moir\'{e} cells in the pattern may look alike,
but they are not exactly identical.

The superstructure, which is a periodic lattice of supercells, does not
suffer from this shortcoming. Unfortunately, it has its own deficiencies.
Namely, the superstructure is defined for commensurate angles $\theta$
only. The period
$L_{\rm sc}$
is not a smooth function of $\theta$: two commensurate angles, $\theta$ and
$\theta'$,
$\theta \approx \theta'$,
may correspond to two very dissimilar
$L_{\rm sc}$.
The existence of two length scales,
$L_{\rm M}$
and
$L_{\rm sc}$,
in tBLG affects its electronic
properties~\cite{PankratovPRB2013}.
While some physical quantities (for example, renormalized Fermi velocity)
are insensitive to sharp variations of
$L_{\rm sc}$
versus $\theta$, others (for example, the gap) are
not~\cite{ourTBLG}.
Consequently, Fermi velocity calculations at commensurate angles are
sufficient for adequate theoretical description; yet, the situation with
the gap is more delicate, as we will show below.

\section{Low-energy effective model}
%%%%%%%%%%%%%%%%%%%%%%%%%%%%%%%%%%%%%%%%%%%%%%%%%%
\label{sect::prelim}
%%%%%%%%%%%%%%%%%%%%%%%%%%%%%%%%%%%%%%%%%%%%%%%%%%

The opening of the gap can be heuristically deduced from the discussion of
Sec.~\ref{sect::geometry}.
Indeed, the low-energy dispersion of the tBLG is characterized by four
Dirac points, two from each layer. At commensurate angles the four points
can be grouped into two equivalence classes, see
Eq.~(\ref{eq::class1})
and~(\ref{eq::class2}).
In other words, while in the original reciprocal space of two sheets of the
single-layer graphene all four Dirac points have different coordinates,
after folding to the first Brillouin zone of the superlattice the
equivalent Dirac points end up in identical locations. The electron states
near equivalent points may be connected by non-zero matrix elements of the
interlayer tunneling operator
$t_\perp$:
in the presence of the superlattice such matrix elements are consistent
with the quasimomentum conservation law. Although the absolute values of
these matrix elements are small, the kinetic energy of electrons near the
Dirac points is small as well. As a result, the interlayer tunneling
qualitatively affects the low-energy spectrum.

To formalize this reasoning, a low-energy effective model is very useful.
In the case of commensurate structures, we can write the low-energy
Hamiltonian in a given corner of the Brillouin zone in the form
\begin{eqnarray}
%%%%%%%%%%%%%%%%%%%%%%%%%%%%%%%%%%%%%%%%%%%%%%%%%%
\label{eq::4x4_ham}
%%%%%%%%%%%%%%%%%%%%%%%%%%%%%%%%%%%%%%%%%%%%%%%%%%
H_{\bf k}^{\rm tBLG}
=
\left(
	\begin{matrix}
		H^{\rm D}_{\gamma \bf k} (0) & M \cr
		M^\dag & H^{\rm D}_{\gamma' \bf k} (\theta) \cr
	\end{matrix}
\right).
\end{eqnarray}
In this expression the (quasi)momentum
${\bf k}$
is measured from the superlattice Brillouin zone corner, while the
single-layer Dirac Hamiltonian
$H^{\rm D}_{\gamma \bf k} (\theta)$
for the rotation angle $\theta$ and cone chirality index
$\gamma = {\bf K, K}'$
equals
\begin{eqnarray}
%%%%%%%%%%%%%%%%%%%%%%%%%%%%%%%%%%%%%%%%%%%%%%%%%%
\label{eq::Dirac_ham}
%%%%%%%%%%%%%%%%%%%%%%%%%%%%%%%%%%%%%%%%%%%%%%%%%%
H^{\rm D}_{\gamma \bf k} (\theta)
=
v_{\text{F}}\, (k_x \sigma_y^\theta \mp k_y \sigma_x^\theta)\,.
\end{eqnarray}
Here
$v_{\text{F}}$
is the Fermi velocity and
$\sigma_{x,y}^\theta = e^{\frac{i \theta}{2}\sigma_z} \sigma_{x,y} e^{-\frac{i \theta}{2}\sigma_z}$
are the ``rotated''\ Pauli matrices, and the sign in
Eq.~(\ref{eq::Dirac_ham})
depends on the chirality index $\gamma$. For structures
$r \ne 3n$,
the chirality indices in the
Hamiltonian~(\ref{eq::4x4_ham})
are unequal
$\gamma \ne \gamma'$.
Otherwise,
$\gamma = \gamma'$.
 The matrix elements
$M_{\alpha\beta}$
of the
$2\times2$
matrix $M$ are given by the equation
\begin{eqnarray}
%%%%%%%%%%%%%%%%%%%%%%%%%%%%%%%%%%%%%%%%%%%%%%%%%%
\label{eq::matrix_M}
%%%%%%%%%%%%%%%%%%%%%%%%%%%%%%%%%%%%%%%%%%%%%%%%%%
M_{\alpha\beta}
=
\sum_{\mathbf{nm}}
	(\psi^{1\alpha}_{\gamma} ({\bf r}^{1 \alpha}_{\bf n}))^*\,\,
	\psi^{2\beta}_{\gamma'} ({\bf r}^{2 \beta}_{\bf m})\,\,
	t_\perp ({\bf r}^{1 \alpha}_{\bf n}, {\bf r}^{2 \beta}_{\bf m})\,.
\end{eqnarray}
In this expression, the interlayer tunneling amplitude
$t_\perp ({\bf r}^{1 \alpha}_{\bf n},{\bf r}^{2 \beta}_{\bf m})$
depends on the location
${\bf r}^{1 \alpha}_{\bf n}$
of an atom in layer~1, sublattice $\alpha$, and the location
${\bf r}^{2\beta}_{\bf m}$
of an atom in layer~2, sublattice $\beta$, see
Eqs.~(\ref{eq::lattice1})
and~(\ref{eq::lattice2}).
The symbol
$\psi^{i\alpha}_{\gamma}$
denotes a spinor component of the wave function in layer
$i=1,2$,
on the sublattice
$\alpha = A,B$
with chirality $\gamma$. The wave function corresponds to the Dirac
point:
$\psi^{1 \alpha}_\gamma  ({\bf r}^{1 \beta}_{\bf n})$
vanishes, if
$\alpha \ne \beta$,
and
$\psi^{1 \alpha}_\gamma  ({\bf r})
\propto
\exp ( \pm i {\bf K}  {\bf r} )$,
where the sign depends on $\gamma$. For layer~2 the wave function is
derived from
$\psi^{1 \alpha}_\gamma  ({\bf r}^{1 \beta}_{\bf n})$
by suitable rotation of the atoms positions.

Strictly speaking, the effective
Hamiltonian~\eqref{eq::4x4_ham}
is applicable only for  large twist angles,
$15^{\circ}\lesssim\theta\lesssim45^{\circ}$.
For smaller angles (or for
$\theta\gtrsim45^{\circ}$), the interlayer matrix elements connecting the
electron states with the same chirality $\gamma$ but different momenta
(constrained, of course, by the superlattice quasimomentum conservation law)
become of
importance~\cite{dSPRL,dSPRB}.
Such coupling terms result in the downward renormalization of the Fermi
velocity. We can take this renormalization into account by
replacing
$v_{\text{F}}$
in
Eq.~\eqref{eq::Dirac_ham}
by the angle-dependent function
$v^*_F(\theta)$.

We calculate the matrix elements of $M$ numerically, both for infinite and
finite samples, with different values of $\theta$. For the latter case, the
twist angle can be arbitrary, not necessarily commensurate. Calculating
$M$ we used the parametrization for the hopping amplitudes
$t_\perp ({\bf r}^{1 \alpha}_{\bf n},{\bf r}^{2 \beta}_{\bf m})$
proposed in
Ref.~\onlinecite{Tang}.
The same parametrization was used in our previous work
Ref.~\onlinecite{ourTBLG}.
Details of the computational procedure are presented in
Appendix~\ref{app::numerical_details}.
Our numerical analysis, as well as arguments of
Ref.~\onlinecite{MelePRB1},
reveals that the matrix $M$ is sensitive to whether the parameter $r$ is a
multiple of $3$, or not.
More precisely, the structure of the matrix $M$ is
the following:
\begin{equation}
%%%%%%%%%%%%%%%%%%%%%%%%%%%%%%%%%%%%%%%%%%%%%%%%%%
\label{eq::matrix_M_def1}
%%%%%%%%%%%%%%%%%%%%%%%%%%%%%%%%%%%%%%%%%%%%%%%%%%
M=\left(\begin{matrix}
0 &me^{i\alpha} \cr
me^{i\beta} & 0 \cr
\end{matrix}\right)\,,
\quad
\text{when\ \ }r\neq 3n\,,
\end{equation}
or
\begin{equation}
%%%%%%%%%%%%%%%%%%%%%%%%%%%%%%%%%%%%%%%%%%%%%%%%%%
\label{eq::matrix_M_def2}
%%%%%%%%%%%%%%%%%%%%%%%%%%%%%%%%%%%%%%%%%%%%%%%%%%
M=\left(\begin{matrix}
0 & 0 \cr
me^{i\beta} & 0\cr
\end{matrix}\right)\,,
\quad
\text{when\ \ }r=3n\,,
\end{equation}
where $m$, $\alpha$, and $\beta$ are real numbers.

\begin{figure}[t]
\centering
\includegraphics[width=0.49\columnwidth]{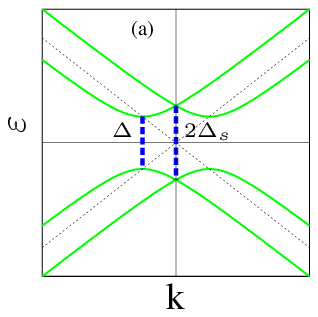}
\includegraphics[width=0.49\columnwidth]{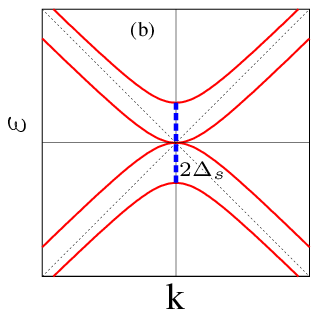}
\caption{
Schematic structure of the low-energy dispersion of twisted bilayer
graphene for
$r \ne 3n$
[panel~(a)] and
$r = 3n$
structures [panel~(b)]. Dotted lines represent two degenerate Dirac cones.
When the matrix $M$
is non-zero, this degeneracy is lifted. The resultant dispersion is shown
by solid [green (a) and red (b)] lines. Vertical dashed lines mark the
energy scales $\Delta$ and
$\Delta_{s}$.
The
$r=3n$
structures have no gap, however, their density of states decreases below
$\Delta_s$.
For
$r \ne 3n$
structures, the spectral gap $\Delta$ and the scale
$2 \Delta_{s}$
are not identical. However, numerical
evidence~\cite{ourTBLG}
suggests that the latter scales are of the same order.
%%%%%%%%%%%%%%%%%%%%%%%%%%%%%%%%%%%%%%%%%%%%%%%%%%
\label{fig::scheme}
%%%%%%%%%%%%%%%%%%%%%%%%%%%%%%%%%%%%%%%%%%%%%%%%%%
}
\end{figure}

The general structure of the
Hamiltonian~\eqref{eq::4x4_ham}
coincides to that proposed in
Ref.~\onlinecite{MelePRB1}.
The main difference lies in the parametrization of the interlayer hopping
amplitudes used to calculate $M$. Our parametrization is able to correctly
describe the limiting case of the AB bilayer
($\theta=0$), as it is explained in Ref.~\onlinecite{ourTBLG}.

The low-energy spectrum is found by diagonalizing the
$4\times4$
matrix
Eq.~(\ref{eq::4x4_ham}).
It consists of four bands with dispersions
$E^{(s)}_{\mathbf{k}}$
($s=1,2,3,4$)
given by
\begin{eqnarray}
%%%%%%%%%%%%%%%%%%%%%%%%%%%%%%%%%%%%%%%%%%%%%%%%%%
\label{Efit}
%%%%%%%%%%%%%%%%%%%%%%%%%%%%%%%%%%%%%%%%%%%%%%%%%%
E^{(1,2,3,4)}_{\mathbf{k}}
\!=
\pm\sqrt{\Delta^2+v_{\text{F}}^2(|\mathbf{k}|\mp k_0)^2}\,,
\quad
\text{if\ \ }
r\neq3n,
\end{eqnarray}
or, for
$r=3n$,
\begin{eqnarray}
%%%%%%%%%%%%%%%%%%%%%%%%%%%%%%%%%%%%%%%%%%%%%%%%%%
\label{EfitReq3}
%%%%%%%%%%%%%%%%%%%%%%%%%%%%%%%%%%%%%%%%%%%%%%%%%%
E^{(1,4)}_{\mathbf{k}}\!=\mp\sqrt{\Delta_s^2+v_{\text{F}}^2\mathbf{k}^2},\;
\\
E^{(2,3)}_{\mathbf{k}}\!
=
\pm \left(
	\sqrt{\Delta_s^2+v_{\text{F}}^2\mathbf{k}^2}-\Delta_s
\right),
\end{eqnarray}
where
\begin{eqnarray}
%%%%%%%%%%%%%%%%%%%%%%%%%%%%%%%%%%%%%%%%%%%%%%%%%%
\label{eq::splitting}
%%%%%%%%%%%%%%%%%%%%%%%%%%%%%%%%%%%%%%%%%%%%%%%%%%
\Delta_s=|m|,
\quad
\Delta=|m\cos[(\alpha-\beta)/2]|,
\\
k_0=m\sin[(\alpha-\beta)/2].
\end{eqnarray}
The
spectra~\eqref{Efit}
and~\eqref{EfitReq3}
are schematically shown in
Fig.~\ref{fig::scheme}.
For structures with
$r\ne 3n$
[see
Fig.~\ref{fig::scheme}(a)],
the tBLG is an insulator with a well-defined gap $\Delta$. If
$r=3n$,
the density of states
$\rho(\varepsilon)$
is finite even at
$\varepsilon = 0$.
However,
$\rho(\varepsilon)$
experiences a depression when
$|\varepsilon|<\Delta_{s}=|m|$,
see
Fig.~\ref{fig::scheme}(b).

\begin{figure}[t]
\centering
\includegraphics[width=1.0\columnwidth]{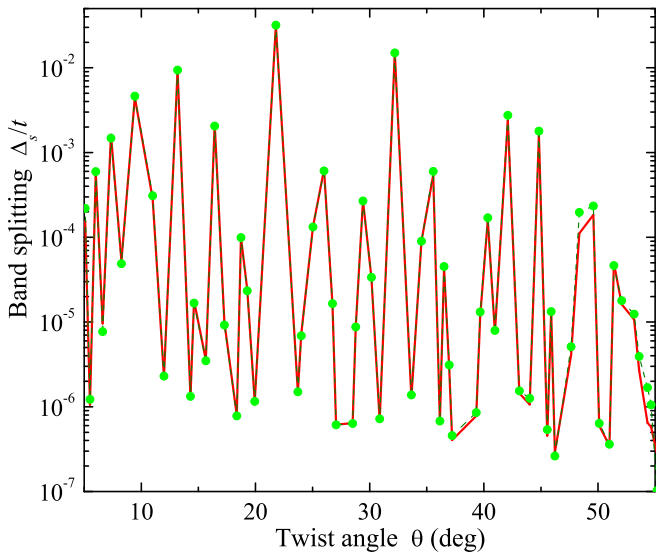}
\caption{Band splitting
$\Delta_{\rm s}$
for ideal superlattices with supercell sizes
$N < 2000$.
Circles (green) connected by the dashed (blue) line present the results of
calculations using
Eq.~\eqref{eq::matrix_M},
while solid (red) line corresponds to the tight-binding calculations of
Ref.~\onlinecite{ourTBLG}.
The data are shown for
$r\neq3n$
structures only. The sharp exponential jumps of
$\Delta_s$,
which we observe in this figure, can exist only in the idealized infinite
tBLG lattice.
}
%%%%%%%%%%%%%%%%%%%%%%%%%%%%%%%%%%%%%%%%%%%%%%%%%%
\label{fig::matrix_elem_supercell}
%%%%%%%%%%%%%%%%%%%%%%%%%%%%%%%%%%%%%%%%%%%%%%%%%%
\end{figure}

The energy scale
$\Delta_s$
will be referred to as the band splitting. We measure here the value of
$\Delta_s$
in units of the graphene's nearest-neighbor hopping amplitude $t$, which is
related to the Fermi velocity
as~\cite{CastroNetoRMP09,ourReview}
$v_{\text{F}}=3ta_0/2$.
Thus, according to the low-energy
model~\eqref{eq::4x4_ham},
the band splitting
$\Delta_s$
is simply a matrix element, whose calculation does not require
diagonalization of any matrix. To check the validity of the
model~\eqref{eq::4x4_ham}
itself we compare
$\Delta_s$
with the results of the tight-binding calculations of the same quantity,
performed in
Ref.~\onlinecite{ourTBLG}.
The curves presented in
Fig.~\ref{fig::scheme}
show a very good correlation between results given by two theoretical
approaches even for small twist angles where the effective
model~\eqref{eq::4x4_ham}
is not formally applicable. For structures with
$r\neq3n$,
the value of
$2\Delta_s$
is larger than the band gap by a factor of order
unity~\cite{ourTBLG}.
Thus, the band splitting given by the modulus of the non-zero matrix
elements in $M$ is a computationally efficient quantity, which can be used
to estimate the possible size of the single-electron gap. In this paper we
will consider the band splitting as a measure of the low-energy spectrum
rearrangement, induced by the interlayer tunneling.

Working with
$\Delta_s$
instead of $\Delta$ reduces the computational complexity. However, the main
issue remains: the elements in the matrix $M$, when calculated for an
infinite superlattice, are not smooth functions of $\theta$, as shown in
Fig.~\ref{fig::matrix_elem_supercell}.
This problem disappears for finite tBLG samples: by construction [see
Eq.~(\ref{eq::matrix_M})],
the matrix elements become analytical functions of the twist angle.
Physically, the finite linear size of the tBLG cluster may indeed
correspond to finite dimensions of a mesoscopic system, or it may mimic a
finite mean free path of an electron due to scattering by disorder, such as
wrinkles and impurities.

Yet, we must remember that a non-zero $m$ in a finite-size system does not
immediately imply the existence of a non-zero gap. The gap could be
observed experimentally only when $m$ exceeds the dimensional quantization
energy
$\delta \varepsilon = v_{\rm F}/L$,
or the disorder scattering rate
$\Gamma \sim v_{\rm F}/l_{\rm m}$
in a sample with disorder. The requirement
\begin{eqnarray}
%%%%%%%%%%%%%%%%%%%%%%%%%%%%%%%%%%%%%%%%%%%%%%%%%%
\label{eq::gap_regime}
%%%%%%%%%%%%%%%%%%%%%%%%%%%%%%%%%%%%%%%%%%%%%%%%%%
\Delta_s(\theta)> \max ( \delta \varepsilon, \Gamma )
\end{eqnarray}
places significant restrictions on the values of $\theta$, for which the
spectrum is gapped. As this condition is violated, the gap is washed away
by external scattering by disorder or edges. This will be discussed in
Section~\ref{sect::numerics}.

\section{Scattering by linear defects}
%%%%%%%%%%%%%%%%%%%%%%%%%%%%%%%%%%%%%%%%%%%%%%%%%%
\label{sect::scattering}
%%%%%%%%%%%%%%%%%%%%%%%%%%%%%%%%%%%%%%%%%%%%%%%%%%

\begin{figure}[t]
\centering
\includegraphics[width=1.0\columnwidth]{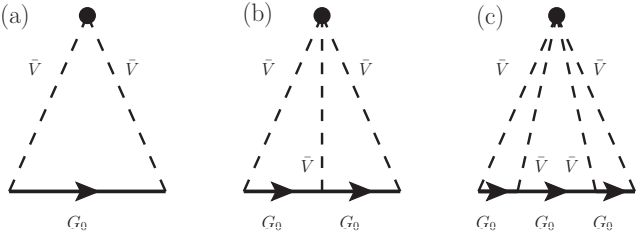}
\caption{Self-energy diagrams for scattering on a single defect. The defect
is represented by a black circle, dashed lines labeled by
${\bar V}$
correspond to the defect potential. Solid lines with arrows are the
electron propagator. Panel~(a) shows the lowest-order contribution to the
self-energy. It equals to
$
{\bar V} \hat{\sigma}_0 = O({\bar V}^2)$
and corresponds to the Born approximation. The higher-order corrections
are shown in panels~(b) and~(c).
%%%%%%%%%%%%%%%%%%%%%%%%%%%%%%%%%%%%%%%%%%%%%%%%%%
\label{fig::wrinkle_bohrn}
%%%%%%%%%%%%%%%%%%%%%%%%%%%%%%%%%%%%%%%%%%%%%%%%%%
}
\end{figure}

We argued in the previous section that disorder can destroy the spectral
gap. In a tBLG there are several possible sources of electron
scattering (electron-electron interaction, point-like neutral and charged
impurities, ``wrinkles'', and others). Studying all of them is beyond the
scope of this paper. In this section, we show that the (inherent for
graphene systems) linear defects (``wrinkles'') are very effective
scatterers in the tBLG, giving rise to a finite mean-free-path
$l_{\rm m}$
when
$\varepsilon\rightarrow 0$.
Our calculations are quite simple, but they allow us to demonstrate the
emergence of the finite energy-independent mean free path in a disordered
system of Dirac electrons.

Let us now consider ``a wrinkle'', a one-dimensional defect stretching along
the $y$-axis. We model this defect by a potential
$V(x,y) = v_{\rm F} {\bar V} \delta(x)$,
where the dimensionless parameter
${\bar V}$
characterizes ``the strength'' of the defect. Neglecting interlayer
hopping, the propagation of the low-energy electron in the graphene layer
is described by the Hamiltonian
Eq.~(\ref{eq::Dirac_ham}).
Within the Born approximation, the self-energy correction due to the
wrinkle equals to
$v_{\rm F} {\bar V} {\hat \sigma}_0/L_x$,
where
$L_x$
is the linear dimension of the sample in the
$x$~direction. The quantity
${\hat \sigma}_0$
is proportional to the usual second-order impurity-scattering loop diagram [see panel~(a) of
Fig.~\ref{fig::wrinkle_bohrn}]
\begin{equation}\label{eq::sgm}
{\hat \sigma}_0 = \frac{v_{\rm F} {\bar V}}{2\pi}
\int dk_x \; G_0 (\varepsilon, {\bf k})\,,
\end{equation}
where the bare Green's function
$G_0$
for the Hamiltonian
Eq.~(\ref{eq::Dirac_ham})
is equal to
\begin{eqnarray}
\nonumber
G_0 =
	\frac{1}{(\varepsilon+i0)^2 - v_{\rm F}^2 |{\bf k}|^2}
	\left(
	\begin{matrix}
		\varepsilon & v_{\rm F} (k_x -i k_y) \cr
		v_{\rm F}( k_x + i k_y ) & \varepsilon \cr
	\end{matrix}
	\right)\,.
\end{eqnarray}
The integral in
Eq.~(\ref{eq::sgm})
is easy to calculate
\begin{eqnarray}
{\hat \sigma}_0
=
-\frac{i{\bar V}}{2 \sqrt{ \varepsilon^2 - v_{\rm F}^2 k_y^2}}
\left(
	\begin{matrix}
		\varepsilon & -i v_{\rm F} k_y \cr
		i v_{\rm F} k_y & \varepsilon \cr
	\end{matrix}
\right)
{\rm sgn}\,\varepsilon\,.
\end{eqnarray}
To obtain the full self-energy it is necessary to sum the self-energy
diagrams to all orders of
${\bar V}$.
The three lowest-order terms of this series are shown in
Fig.~\ref{fig::wrinkle_bohrn}.
Since the $n$-th order diagram is proportional to
${\hat \sigma}_0^n$,
the summation is performed trivially, and one derives
\begin{eqnarray}
{\hat \Sigma}_0
=
\frac{v_{\rm F}}{L_x}
\frac{{\bar V} {\hat \sigma}_0}{1-{\hat \sigma}_0}\,.
\end{eqnarray}
This self-energy conserves the energy $\varepsilon$ and momentum
$k_y$.
As for
$k_x$,
it is not conserved: upon scattering off the wrinkle, the momentum
projection
$k_x$
can change arbitrarily with finite probability. For an ensemble of wrinkles
we must average over the location of the wrinkle. This procedure restores
the conservation of
$k_x$,
and the resultant self-energy becomes
\begin{eqnarray}
%%%%%%%%%%%%%%%%%%%%%%%%%%%%%%%%%%%%%%%%%%%%%%%%%%
\label{eq::wrinkle_se1}
%%%%%%%%%%%%%%%%%%%%%%%%%%%%%%%%%%%%%%%%%%%%%%%%%%
{\hat \Sigma}
=
n_{\rm w}
\frac{v_{\rm F} {\bar V} {\hat \sigma}_0}{1-{\hat \sigma}_0}\,,
\end{eqnarray}
where
$n_{\rm w}$
is the concentration of the ``wrinkles'' (it has a dimension of the inverse
length). The self-energy
${\hat \Sigma}$
is diagonal both in $\varepsilon$ and in
${\bf k}$.

The averaging over the location of the wrinkle, which we performed to
derive
Eq.~(\ref{eq::wrinkle_se1}),
must be supplemented by the averaging over the orientations of the
wrinkles. After all, in a generic situation, an ensemble of wrinkles is
likely to be fairly isotropic. To perform this averaging it is useful to
notice that the matrix
${\hat \sigma}_0$
has two eigenvalues
\begin{eqnarray}
%%%%%%%%%%%%%%%%%%%%%%%%%%%%%%%%%%%%%%%%%%%%%%%%%%
\label{eq::sigma_eig}
%%%%%%%%%%%%%%%%%%%%%%%%%%%%%%%%%%%%%%%%%%%%%%%%%%
\sigma^\pm
=
-\frac{i\bar V}{2\sqrt{\varepsilon^2 - v_{\rm F}^2 k_y^2}}
(\varepsilon \pm v_{\rm F} k_y)\, {\rm sgn}\, \varepsilon\,,
\end{eqnarray}
which correspond to the eigenvectors
$(1,\pm i)/\sqrt{2}$.
The matrix
$\hat \Sigma$
will have the same eigenvectors. The eigenvalues of
$\hat\Sigma$
can be found using
Eqs.~\eqref{eq::wrinkle_se1}
and~\eqref{eq::sigma_eig}.

Since the eigenvectors of
$\hat \Sigma$
are independent of both $\varepsilon$ and
$k_{y}$,
we need to average the eigenvalues only. Further simplification can be
obtained if we work on the mass surface. There one can write
$v_{\rm F} k_y = \varepsilon \sin \phi$,
where $\phi$ denotes the angle of incidence of the electron on the wrinkle.
The eigenvalues of
$\hat \Sigma$
on the mass surface are
\begin{eqnarray}
\Sigma^\pm_{\rm m.s.}
=
-n_{\rm w}
\frac{i{\bar V}^2(1 \pm \sin \phi)}
	{2 |\cos \phi| + i {\bar V} (1 \pm \sin \phi)}\,.
\end{eqnarray}
The required integration over $\phi$ is well-defined for any non-zero
${\bar V}$.
It is clear that after such an integration both eigenvalues become
identical, and the averaged self-energy is proportional to the scalar
matrix. In the limit of small
${\bar V}$
we obtain
\begin{eqnarray}
\Sigma_{\rm m.s.}
=
-i{\bar V}^2 n_{\rm w}
\int\limits_{-\pi/2}^{\pi/2}\!\!
\frac{d \phi}{2\pi}
\frac{ \cos \phi} { \cos^2 \phi + {\bar V}^2}\,,
\end{eqnarray}
which implies that the scattering rate is
\begin{eqnarray}
%%%%%%%%%%%%%%%%%%%%%%%%%%%%%%%%%%%%%%%%%%%%%%%%%%
\label{eq::Gamma_wrinkle}
%%%%%%%%%%%%%%%%%%%%%%%%%%%%%%%%%%%%%%%%%%%%%%%%%%
\Gamma\propto n_{\rm w} {\bar V}^2 \ln{\bar V}\,.
\end{eqnarray}

This relation for the scattering rate was derived under the assumption that
the multiple-wrinkle scattering effects may be neglected. Thus, the
localization cannot be described in the framework of the above procedure.
The expression for $\Gamma$ is energy-independent, and is valid at low
energies. Unlike point-like impurities, whose scattering in graphene
becomes weaker (for weak impurity potentials) as the quasiparticle energy
lowers~\cite{CastroNetoDisorder2008},
the linear defects scatter well even at the Dirac point. Consequently, the
electrons acquire a finite mean free path
$l_{\rm m} \sim v_{\rm F}/\Gamma < \infty$.
This limits the coherent propagation of the electron wave packet, and
destroys weak interference effects due to superstructures with large
supercell sizes.

\section{Gap and band splitting for finite samples}
%%%%%%%%%%%%%%%%%%%%%%%%%%%%%%%%%%%%%%%%%%%%%%%%%%
\label{sect::numerics}
%%%%%%%%%%%%%%%%%%%%%%%%%%%%%%%%%%%%%%%%%%%%%%%%%%

\begin{figure}[t]
\centering
\includegraphics[width=1.0\columnwidth]{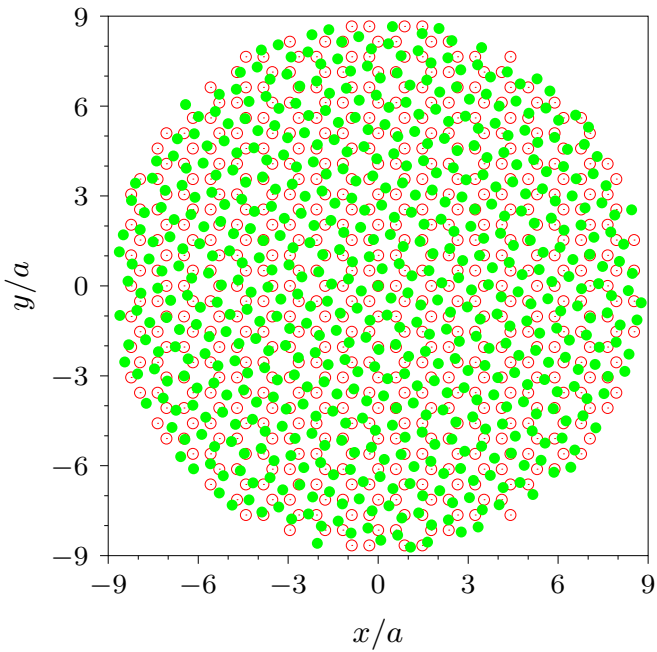}
\caption{Cluster of tBLG. Radius
$R=15a_0=5\sqrt{3}a$,
with a rotation angle
$\theta = 16.7^\circ$.
The bottom layer is shown by open (red) circles, while the top (rotated)
layer by filled (green) circles.
}
%%%%%%%%%%%%%%%%%%%%%%%%%%%%%%%%%%%%%%%%%%%%%%%%%%
\label{fig::cluster}
%%%%%%%%%%%%%%%%%%%%%%%%%%%%%%%%%%%%%%%%%%%%%%%%%%
\end{figure}

Thus, the coherent propagation of an electron in a tBLG sample is always
limited to some finite length scale. In the present study, to mimic this
length we modeled a tBLG as a cluster of finite size, see
Fig.~\ref{fig::cluster}.
The cluster has circular shape, it consists of the sites of the tBLG
lattice whose distance from the origin is less than the cluster radius $R$.
For example, the cluster in
Fig.~\ref{fig::cluster}
has
$R=15a_0=5\sqrt{3}a$.

\begin{figure}[t!]
\centering
\includegraphics[width=1.0\columnwidth]{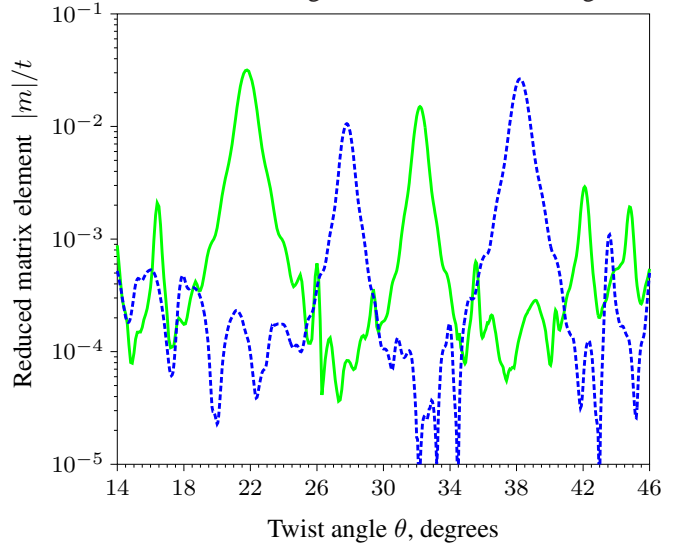}
\caption{The matrix element
$|m|$
as a function of the twist angle $\theta$. The radius of the cluster is
$R=60a_0$.
The (green) solid curve shows the value of
$|m|$
which is responsible for the opening of the band gap $\Delta$ in the
structures with
$r \neq 3n$,
Figure~\ref{fig::scheme}(a).
The (blue) dashed curve shows the matrix element inducing the band
splitting in structures with
$r=3n$,
Figure~\ref{fig::scheme}(b).
The maxima of both curves are located at the angles
$\theta_{(m_0,r)}$
corresponding to the superstructures with small $r$ and
$m_0$.
For example, the strongest maxima of the (green) solid curve are at
$\theta_{(1,1)} \approx 21.8^\circ$
and
$\theta_{(1,2)}= 32.2^\circ$.
For the (blue) dashed curve these are at
$\theta_{(2,3)} = 27.8^\circ$
and
$\theta_{(1,3)} = 38.2^\circ$.
Note that
$\theta_{(1,1)} +
\theta_{(1,3)} = 60^\circ$
and
$\theta_{(1,2)} + \theta_{(2,3)} =
60^\circ$,
in agreement with
Eq.~(\ref{eq::conj_struct}).
%%%%%%%%%%%%%%%%%%%%%%%%%%%%%%%%%%%%%%%%%%%%%%%%%%
\label{fig::all_data}
%%%%%%%%%%%%%%%%%%%%%%%%%%%%%%%%%%%%%%%%%%%%%%%%%%
}
\end{figure}

As shown in
Section~\ref{sect::prelim}
in the framework of the low-energy
model~\eqref{eq::4x4_ham},
the band splitting
$\Delta_s$
is equal to the modulus of the non-zero matrix element(s) of the matrix
$M$, see
Eq.~\eqref{eq::splitting}.
Likewise, the band gap $\Delta$ is proportional to
$|m|$.
We calculate these matrix elements numerically as prescribed by
Eqs.~(\ref{eq::matrix_M}),~(\ref{eq::matrix_M_def1}),
and~(\ref{eq::matrix_M_def2}),
for a range of $R$'s and $\theta$'s (additional technical details can be
found in the Appendix). The typical behavior of
$|m|$
is shown in
Fig.~\ref{fig::all_data},
where numerical data, in the window
$14^\circ < \theta < 46^\circ$,
is plotted for a cluster of radius
$R/a_0 = 60$.
Both
$r\neq 3n$
and
$r=3n$
data are presented. The pronounced peaks in
Fig.~\ref{fig::all_data}
occur at ``good'' angles corresponding to the superlattices with small
supercells. Smaller peaks may be associated with some finite-size effects:
these peaks sharply weaken when $R$ is increased.

It is
known~\cite{ourReview}
that for a
$r\neq3n$
structure, characterized by the twist angle $\theta$, one can construct a
conjugate
$r=3n$
structure with the angle
\begin{eqnarray}
%%%%%%%%%%%%%%%%%%%%%%%%%%%%%%%%%%%%%%%%%%%%%%%%%%
\label{eq::conj_struct}
%%%%%%%%%%%%%%%%%%%%%%%%%%%%%%%%%%%%%%%%%%%%%%%%%%
\theta' = 60^\circ - \theta\,,
\end{eqnarray}
such that both structures have the same supercell size. The data in
Fig.~\ref{fig::all_data}
illustrates this relation: two strongest peaks are located at angles
$21.7^\circ$
and
$38.2^\circ$,
whose sum equals to
$60^\circ$.
The same is true for the pair of the second-strongest peaks at
$27.8^\circ$
and
$32.2^\circ$.

\begin{widetext}
\begin{figure*}
\centering
\includegraphics[width=1.0\textwidth]{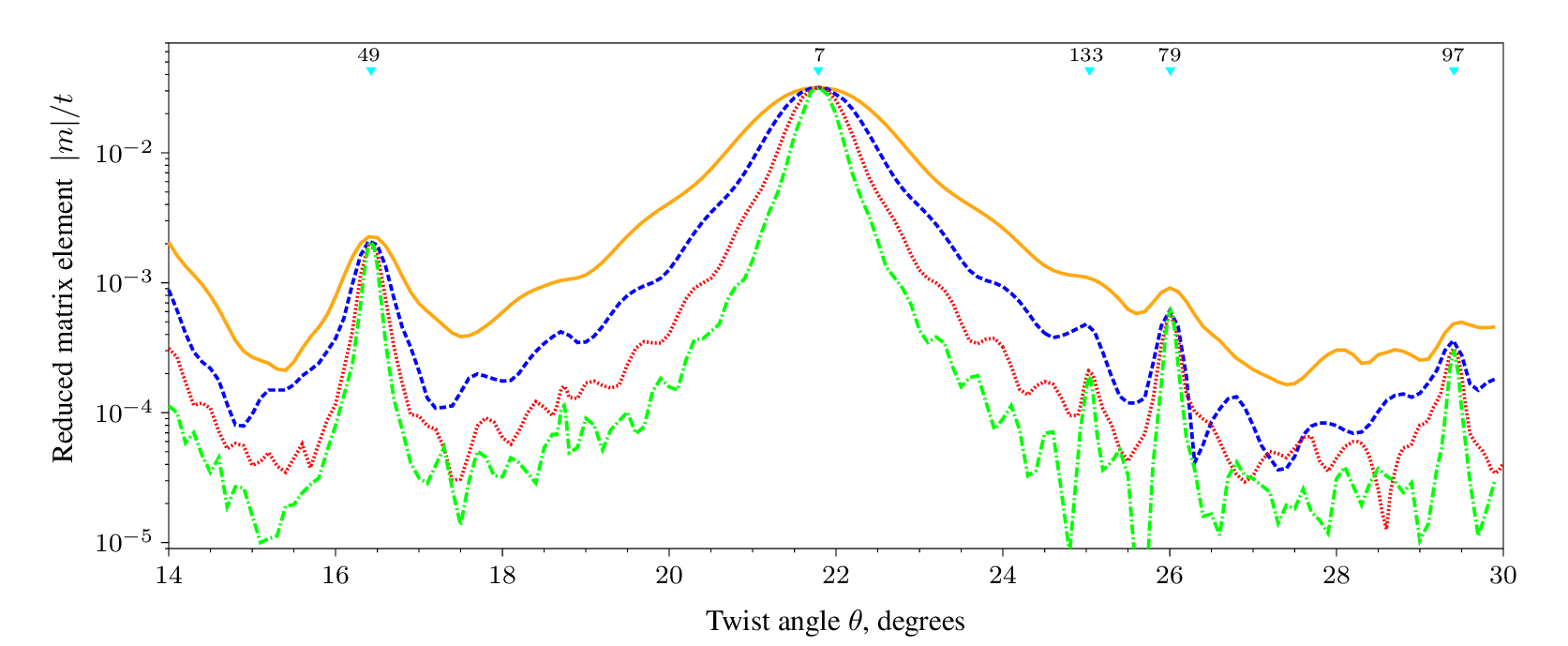}
\caption{The matrix element
$|m|$
as a function of the twist angle for clusters of different radii. Four
curves in this figure correspond to the following values of the effective
cluster radius (for details, see Appendix):
$R=40a_0$
is shown by the solid (yellow) curve on top,
$R=60a_0$
by the dashed (blue) curve,
$R=90a_0$
by the dotted (red) curve, and
$R=130a_0$
is shown by the dash-dotted (green) curve at the bottom. The triangles at
the top edge of the figure mark the commensurate angles with relatively
small supercell linear size
($L_{\rm sc} \leq 20a_0$).
The numbers above these triangles show the number of graphene's unit cells inside the supercell ($N/4$).
One can notice that, at sufficiently large values of $R$, a peak forms at every
marked angle.
}
%%%%%%%%%%%%%%%%%%%%%%%%%%%%%%%%%%%%%%%%%%%%%%%%%%
\label{fig::gap}
%%%%%%%%%%%%%%%%%%%%%%%%%%%%%%%%%%%%%%%%%%%%%%%%%%
\end{figure*}
\end{widetext}

The matrix element
$|m|$,
responsible for the band gap in the spectrum of
$r \ne 3n$
superstructures, is plotted for clusters of different sizes in
Fig.~\ref{fig::gap}.
We see that for a generic value of the twist angle, the quantity
$|m|$
quickly decreases with increasing $R$. At the same time, when $\theta$
corresponds to commensurate superlattices with small supercell size,
$|m|$
remains constant
($\theta \approx 16.7^\circ$,
$21.8^\circ$). For somewhat larger supercell sizes
($\theta = 25.0^\circ$,
$26.0^\circ$,
$29.4^\circ$)
the band splitting initially decreases, only to saturate at larger radii.
The stabilization occurs when $R$ sufficiently exceeds the supercell size.
As an example, consider the
$\theta = 26.0^\circ$
and
$\theta = 29.4^\circ$
twist angles. In both cases, the matrix element stops changing when
$R \geq 60a_0$.
To weaken the edge effects for a finite cluster, our numerical procedure
(see Appendix for details) confines the electron wave function within the
effective radius
$R_{\rm eff} < R$,
defined as
\begin{eqnarray}
%%%%%%%%%%%%%%%%%%%%%%%%%%%%%%%%%%%%%%%%%%%%%%%%%%
\label{eq::r_eff}
%%%%%%%%%%%%%%%%%%%%%%%%%%%%%%%%%%%%%%%%%%%%%%%%%%
R_{\rm eff} \approx R/2.2.
\end{eqnarray}
A physical cluster radius of
60$a_0$
corresponds to the effective radius
$R_{\rm eff} \approx 27a_0$.
The latter number is comparable to the supercell size of
15$a_0$
and
16$a_0$
for such values of $\theta$. If
$\theta = 25.0^\circ$,
the growth of
$|m|$
is stabilized at
$R=90a_0$,
or
$R_{\rm eff} = 41a_0$.
This is of the order of
$L_{\rm sc} = 20a_0$
for the
$\theta = 25.0^\circ$
superstructure. We see that for these three angles the matrix element
saturates when
$R_{\rm eff} \gtrsim 2 L_{\rm sc}$.

The curves shown in
Fig.~\ref{fig::gap}
demonstrate that for finite clusters the matrix elements responsible for
the gap are smooth functions of $\theta$, unlike the data for infinite
systems shown in
Fig.~\ref{fig::matrix_elem_supercell}.
However, the results presented in
Fig.~\ref{fig::gap}
should not be interpreted as the dependence of the band gap versus the
twist angle. As
condition~(\ref{eq::gap_regime})
implies, to decide if the tBLG spectrum has a gap (more precisely,
pseudogap), it is necessary to compare
$|m|$
against the dimensional quantization energy
\begin{eqnarray}
\delta \varepsilon
\sim
\frac{v_{\rm F}}{R}\,.
\end{eqnarray}
Equivalently, the scale
$v_{\rm F}/ |m|$
should be smaller than $R$.

\begin{figure}%[btp]
\centering
\includegraphics[width=0.85\columnwidth]{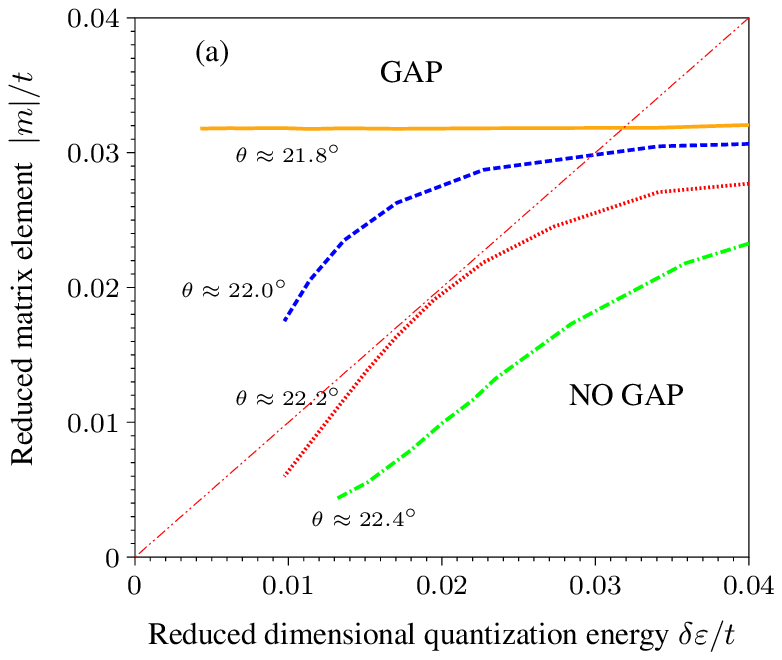}%\vspace{0.5cm}
\vspace{0.5cm}
\includegraphics[width=0.85\columnwidth]{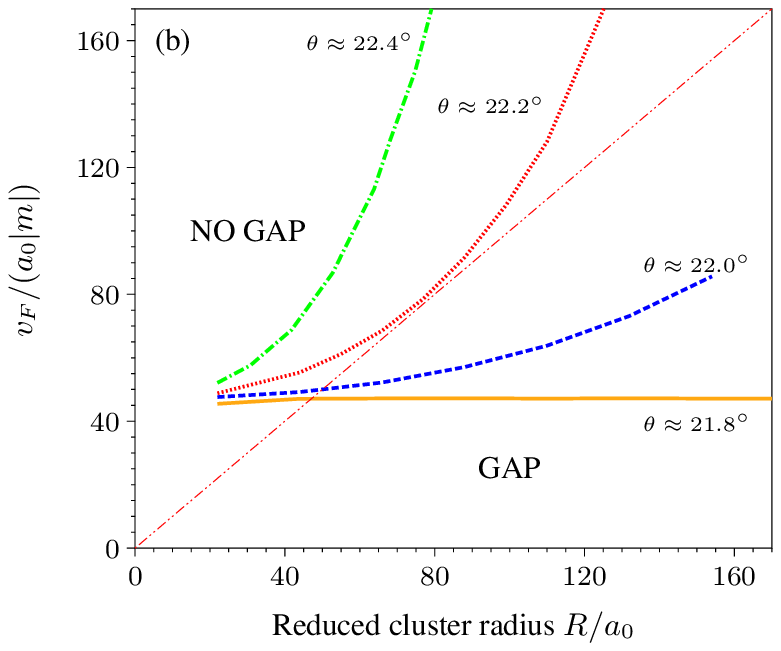}
\caption{Crossover between the gapless and gapped regimes. Panel~(a) shows
the parametric plots of the matrix element
$|m(R)|$
versus the dimensional quantization energy
$\delta \varepsilon (R)$
for several values of $\theta$. The twist angle $\theta$ is constrained to
the vicinity of the ``good" commensurate angle
$\theta_0 \approx 21.8^\circ$.
The thin dash-dotted (red) straight line is determined by the equation
$|m| = \delta \varepsilon$,
marking the crossover between gapless
($|m| < \delta \varepsilon$)
and gapped
($|m| > \delta \varepsilon$)
spectra. Exactly at the commensurate angle [solid (yellow) curve] the system
is gapless at larger
$\delta \varepsilon$
(smaller $R$). It enters into a gapped regime for larger cluster size
(smaller
$\delta \varepsilon$).
The [dashed (blue)] curve for
$22.0^\circ$
demonstrates similar behavior. When deviation from the ``good'' angle is
higher [e.g.,
$\theta \approx 22.4^\circ$,
dash-dotted (green) curve] the system never enters into the gapped regime.
The angle
$\theta^* \approx 22.2^\circ$
separates two types of behavior [and the corresponding dotted (red) curve touches
the line
$|m| = \delta \varepsilon$
when
$\delta \varepsilon \approx 0.015 t$].
In panel~(b) the same data are plotted in a different manner: instead of
comparing the dimensional quantization energy and
$|m|$,
panel~(b) allows us to compare the cluster radius $R$ and the length scale
$v_{\rm F}/|m|$.
The results for
$\theta<21.8^\circ$
are almost symmetric.
%%%%%%%%%%%%%%%%%%%%%%%%%%%%%%%%%%%%%%%%%%%%%%%%%%
\label{fig::xover_angle}
%%%%%%%%%%%%%%%%%%%%%%%%%%%%%%%%%%%%%%%%%%%%%%%%%%
}
\end{figure}

To describe the crossover between gapless and gapped regimes, let us
analyze
Fig.~\ref{fig::xover_angle},
where we replotted the data presented in
Fig.~\ref{fig::gap}
in a new manner: for a given curve, the angle $\theta$ is fixed, while the
cluster size varies. The range of the twist angles in
Fig.~\ref{fig::xover_angle}
is restricted to the vicinity of
$\theta_0 \approx 21.8^\circ$.
We consider here only the angles
$\theta < \theta_0$,
since for
$\theta > \theta_0$
the results are almost symmetric. The angle
$\theta_0$
corresponds to the smallest supercell possible for a tBLG. At
$\theta=\theta_0$,
the value of
$|m|$
is the largest, see
Fig.~\ref{fig::all_data}.

Panel~(a) of
Fig.~\ref{fig::xover_angle}
shows
$|m(R)|$
as an implicit function of the dimensional quantization energy
$\delta \varepsilon (R)$.
In panel~(b) the length scale
$v_{\rm F} / |m|$
is plotted as a function of $R$. In both panels of
Fig.~\ref{fig::xover_angle}
the dash-dotted straight lines are set by the equation
$|m| = \delta \varepsilon$.
These lines mark the crossover from the gapless
($|m| < \delta \varepsilon$)
to the gapped
($|m| > \delta \varepsilon$)
regimes.

The crossover can occur when the size of the cluster becomes sufficiently
large. For example, if the twist angle is exactly commensurate (solid green
curves on both panels), the increase of $R$, and concomitant decrease of
$\delta \varepsilon$,
pushes the sample from a gapless state to a state with single-electron gap.
The data presented suggest that the crossover occurs
when
$R \approx 50 a_0$,
or,
equivalently,
$R_{\rm eff} \approx 23 a_0$.

If deviations from the commensurate angle is small
($\theta \approx 22.0^\circ$,
dashed blue curve) the situation remains qualitatively the same: the
gapless regime at small $R$ is replaced by a gapped regime at larger $R$.
For stronger deviations (e.g.,
$\theta \approx 22.4^\circ$,
dash-dotted orange curve) the system never leaves the gapless regime for
any $R$. When
$\theta = \theta^* \approx 22.2^\circ$,
the corresponding curve touches the crossover line. The angle
$\theta^*$
separates two types of behavior. If
$\theta > \theta^*$,
the system is gapless even when the cluster is large. When
$\theta_0 < \theta < \theta^*$,
the crossover to the gapped regime can occur with increasing $R$. This
analysis demonstrates that, to observe the single-electron gap caused by
the interlayer tunneling near the commensurate angle
$21.8^\circ$,
the twist must be controlled with an accuracy
$\delta \theta \approx |\theta^* - \theta_0| \approx 0.4^\circ$.

The same procedure can be performed near another ``good'' angle
$\theta\approx 32.2^\circ$,
corresponding to
$r=2$
and
$m_0=1$,
see
Fig.~\ref{fig::all_data}.
The matrix element for this superstructure is roughly two times smaller
than that for the structure with
$r=m_0=1$
($\theta\approx 21.8^\circ$).
Consequently, the radius of the clusters must be doubled to have a chance
to be in the gapped regime. The increase in $R$ translates into a more
stringent requirement on the fine-tuning of $\theta$: to observe the gap,
the deviation from the commensurate angle must satisfy
$\delta \theta \sim 0.1^\circ$.
Such a decrease in the allowed deviation of
$\delta \theta$
can be understood as follows. A smaller
$|m|$
implies that a larger $R$ is necessary to enter the gapped regime. However,
for larger clusters the maxima in
Fig.~\ref{fig::gap}
become sharper; consequently, the matrix element becomes very sensitive to
the value of the twist angle. Therefore, even a weak deviation from the
``good'' angle may push
$|m|$
below
$\delta \varepsilon$.

Investigations of superstructures with larger supercells place heavy
requirements on computational resources. Indeed, large supercells
correspond to exponentially small matrix elements, which means that
exponentially large cluster sizes must be studied to enter the regime
$\delta \varepsilon > |m|$.
Such studies are computationally impractical. Thus, we must rely on the
information collected above to draw conclusions.

\section{Discussion and conclusions}
%%%%%%%%%%%%%%%%%%%%%%%%%%%%%%%%%%%%%%%%%%%%%%%%%%
\label{sect::disc_and_concl}
%%%%%%%%%%%%%%%%%%%%%%%%%%%%%%%%%%%%%%%%%%%%%%%%%%

The single-electron gap in the tBLG spectrum is a particularly challenging
and interesting property. This gap demonstrates ``fractal'' oscillations
when changing the twist angle (shown in
Fig.~\ref{fig::all_data}),
unlike, for example, the Fermi velocity, which varies smoothly. These
oscillations are an artifact of the assumption that an electron propagates
inside a perfect infinite tBLG lattice. In a realistic situation, the
coherent propagation of a wave packet through the lattice is limited by the
finiteness of the sample size $L$, and/or disorder scattering.

A particular example of disorder, one-dimensional wrinkles, was considered
in
Sec.~\ref{sect::scattering}.
Defects of this kind are of interest due to two main reasons. First, it is
an inherent type of disorder in graphene systems. Second, a linear defect
is an effective source of scattering for low-energy Dirac quasiparticles,
which is of importance for tBLG, with its flat bands and low-energy Van
Hove singularity. Let us also comment that, since one-dimensional defects
are very effective in destroying coherence, the fragile phenomenology of
the marginal Fermi liquid, predicted for undoped
graphene~\cite{guinea_marginal,DasSarmaMarginal,GrapheneInteractionReview},
may not survive in a sample with a sufficient concentration of wrinkles.

When the coherent propagation length
$l_{\rm coh}={\rm min}\{l_{\rm m}, L \}$
is finite, the diffraction effects associated with the superstructures with
large supercells are destroyed. As a result, small gaps corresponding to
such superlattices disappear. The stronger gaps can become observable,
provided that (a) the length
$l_{\rm coh}$
is sufficiently large, and (b) the deviation of the twist angle from a
``good'' value is sufficiently small.

The condition~(a) is very general. It is necessary to remember that the
band splitting
$\Delta_s$
and, consequently, the gap is washed away by the disorder, or masked by
finite size quantization, if
$\Delta_s < v_{\rm F}/ l_{\rm coh}$.
This implies that the gap, or pseudogap, may be observed only when
$l_{\rm coh} \gg v_{\rm F}/\Delta_s$.

Regarding condition~(b), we have seen that the matrix element responsible
for the opening of the gap is very sensitive to the shift
$\delta \theta$
of the twist angle away from the ``good'' value. If $\theta$ coincides with
a ``good'' angle
($\delta \theta = 0$),
the matrix element becomes independent of
$l_{\rm coh}$
for sufficiently large
$l_{\rm coh}$.
Thus, exactly at a ``good'' angle the pseudogap or gap can be measured in a
large sample of high purity. For small deviations from such an angle, the
value of
$\Delta_s$
decreases somewhat as
$l_{\rm coh}$
grows, but the same qualitative picture endures.

\begin{figure}[t]
\centering
\includegraphics[width=1\columnwidth]{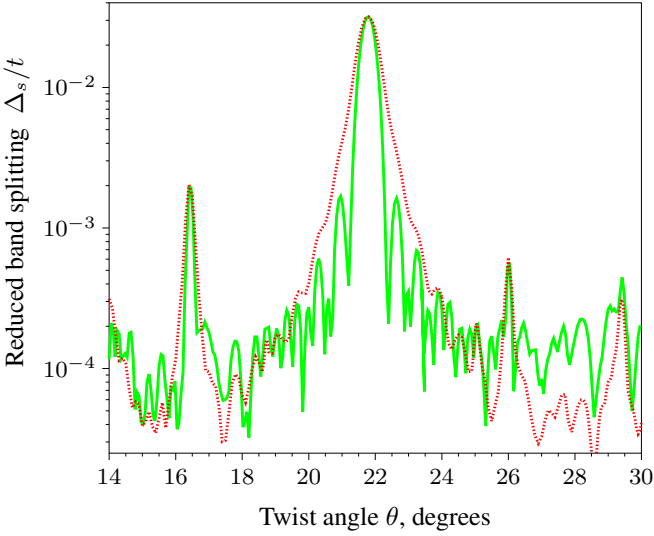}
\caption{
Effect of the exponential decay of the wave function. The band splitting
calculated for different values of $\zeta$, see
Eq.~(\ref{eq::app::wave_func}).
The
$\zeta=2.2$
data (dotted red curve) shows a smoother behavior than the
$\zeta=1.1$
data (solid green curve). The radius of the cluster is 90$a_0$ for both
curves.
}
%%%%%%%%%%%%%%%%%%%%%%%%%%%%%%%%%%%%%%%%%%%%%%%%%%
\label{fig::edge_smooth}
%%%%%%%%%%%%%%%%%%%%%%%%%%%%%%%%%%%%%%%%%%%%%%%%%%
\end{figure}

However, as
$\delta \theta$
departs from zero, the stabilization of the gap and the band splitting
$\Delta_s$
at larger
$l_{\rm coh}$
does not occur, see
Fig.~\ref{fig::xover_angle}.
Instead, the matrix element quickly collapses with increasing
$l_{\rm coh}$.
As a result, for large deviations of $\theta$ from the ``good'' angle, the
gapped regime never occurs.

Our analysis demonstrates that the experimental observation of the
single-electron gap caused by the superlattice scattering is extremely
unlikely, unless a very precise tuning of the twist angle to the ``good''
values is achieved. Such control may be enforced
externally~\cite{Koren2016}.
Alternatively, one can speculate that commensurate angles correspond to
local minima of the interlayer interaction potential. Consequently, the
bilayer might spontaneously lock the twist angle to these angle values.
However, such a possibility is, at this point, nothing but a hypothesis,
and further research is required to support or refute it.

To conclude, we studied the dependence of the single-electron gap in finite
clusters of tBLG. We demonstrated that the variation of the twist angle
causes a crossover between gapless and gapped regimes, provided that the
coherent propagation of an electron is limited by some finite length scale.
Either the finiteness of the sample or the mean free path due to the
disorder scattering may generate the latter length scale. To observe the
gap experimentally it is necessary to have a sample of sufficient purity,
and possess the ability to tune the twist angle accurately.

\appendix

\section{Details of numerical procedure}
%%%%%%%%%%%%%%%%%%%%%%%%%%%%%%%%%%%%%%%%%%%%%%%%%%
\label{app::numerical_details}
%%%%%%%%%%%%%%%%%%%%%%%%%%%%%%%%%%%%%%%%%%%%%%%%%%
Here we briefly outline additional details of our numerical procedure which
were too specialized to be included in the main text.

To calculate the matrix elements in
Eq.~\eqref{eq::matrix_M}
we use the following expression for the inter-layer hopping amplitude
\begin{eqnarray}
&&t_{\bot}(\mathbf{r};\mathbf{r}')=
\cos^2\!\alpha\; V_{\sigma}(\mathbf{r};\mathbf{r}')
+\sin^2\!\alpha\; V_{\pi}(\mathbf{r};\mathbf{r}')\,,\nonumber\\
&&\cos\alpha=\frac{d}{\sqrt{d^2+(\mathbf{r}-\mathbf{r}')^2}}\,,
\end{eqnarray}
where
$d=3.32$\,\AA\ is the interlayer distance,
$\mathbf{r}$
and
$\mathbf{r}'$
are 2D coordinates of the carbon atoms in the bottom and top layers,
respectively, and
$V_{\sigma}$
and
$V_{\pi}$
are the `Slater-Koster' functions, which we choose in the form of Eq.~(1)
of
Ref.~\onlinecite{Tang}.
In that paper the tunneling amplitude of an electron
from one atom to another depends not only on the relative positions these
two atoms, but also on the positions of other atoms in the crystal via the
screening function $S$. The latter one has several fitting parameters,
which we choose such that the function
$t_{\bot}(\mathbf{r};\mathbf{r}')$
would correctly describe the first several interlayer hopping amplitudes of
the AB bilayer
($\theta=0$)
graphene. More details can be found in
Ref.~\onlinecite{ourTBLG}.

It is known~\cite{fujita_edge1996,maximov_edge2013,zagor_edge2015,meso_review}
that various types of localized states exist at the edges of graphene and
graphene-based systems. Since we are interested in the bulk behavior, the
influence of such states is to be reduced as much as possible. To decrease
the effects of edge phenomena we introduced an exponential decay of the
wave function from the cluster center toward the edges. Specifically, the
matrix element
Eq.~(\ref{eq::matrix_M})
is calculated using the wave function for the layer~$1$
\begin{eqnarray}
%%%%%%%%%%%%%%%%%%%%%%%%%%%%%%%%%%%%%%%%%%%%%%%%%%
\label{eq::app::wave_func}
%%%%%%%%%%%%%%%%%%%%%%%%%%%%%%%%%%%%%%%%%%%%%%%%%%
\psi^{1 \alpha}_{\gamma} ({\bf r}^{1 \beta}_{\bf n})
=
{\cal N} \exp(-i{\bf K}_{\gamma} {\bf r}^{1 \alpha}_{\bf n}
- \zeta |{\bf r}^{1 \alpha}_{\bf n}|/R)\delta_{\alpha\beta}\,,
\end{eqnarray}
where
${\bf K}_{\gamma}$
is the Dirac point corresponding to the chirality $\gamma$,
$\delta_{\alpha\beta}$
is the Kronecker symbol, and
$\zeta= 2.2$
is a numerical coefficient. In layer~2 the wave function is constructed in
a similar manner. A wave function in layer~2 matches a wave function in
layer~1 after an appropriate rotation. For finite samples, the wave
functions are normalized to unity, with
${\cal N}$
being the normalization constant. It is worth noting that for infinite
samples a different normalization condition should be used:
$\sum_{\mathbf{n}}
	|\psi^{i\alpha}_{\gamma}({\bf r}^{i\alpha}_{\bf n})|^2=1$,
where the sum is taken over sites inside one supercell.

The magnitude of the wave function decreases away from the cluster center.
The value of the numerical factor
$\zeta= 2.2$
was chosen empirically. If $\zeta$ is too large, the effective size of the
cluster
\begin{eqnarray}
R_{\rm eff} \sim R/\zeta
\end{eqnarray}
shrinks significantly below its nominal radius $R$; thus, we are forced to
study computationally expensive cases of large $R$. If $\zeta$ is too
small, the edge effects make the data very ``noisy'', see
Fig.~\ref{fig::edge_smooth}.

Interpreting our numerical data one must keep in mind that for finite $R$
and arbitrary $\theta$ the absolute values of the non-zero elements of the
matrix $M$,
Eq.~(\ref{eq::matrix_M_def1}),
may be slightly different from each other. However, we checked numerically
that this disparity is not significant, at least for commensurate
structures and larger clusters.

The data presented were collected for clusters in which the rotation axis
passes through the geometrical center of the cluster. One can shift the
rotation axis off the cluster center by the vector
${\bf T} = n {\bf a}_1 + m {\bf a}_2$,
where
$n,m$
are integers. As long as
$|{\bf T}| \ll R$,
it is expected that the matrix $M$ is independent of
${\bf T}$.
We verified that this is indeed the case.

\section*{Acknowledgments.}

This work was supported in part by RFBR (Grants
Nos.~14-02-00276, 14-02-00058, 15-02-02128).
F.N. was partially supported by: the RIKEN iTHES Project,
the MURI Center for Dynamic Magneto-Optics via the AFOSR Award
No.~FA9550-14-1-0040,
the Japan Society for the Promotion of Science (KAKENHI),
the ImPACT program of JST, CREST,
and a grant from the John Templeton Foundation.

\vspace*{-0.1in}

%\bibliographystyle{apsrevlong_no_issn_url}
%\bibliography{twist}

\end{document}